\documentclass[12pt]{article}
\usepackage{amsmath}
\usepackage{graphicx}
\usepackage{natbib}
\usepackage{url} 

\usepackage{bm}
\usepackage{amssymb}
\usepackage[ruled]{algorithm2e}
\usepackage{booktabs}
\usepackage{multicol}
\usepackage{hyperref}
\usepackage[skip=0pt]{caption}

\newcommand{\blind}{0}

\begin{document}

\def\spacingset#1{\renewcommand{\baselinestretch}%
{#1}\small\normalsize} \spacingset{1}


\if0\blind
{
  \title{\bf Model-Based Clustering with Sequential Outlier Identification using the Distribution of Mahalanobis Distances}
  \author{Ult\'{a}n P. Doherty\thanks{
    The authors gratefully acknowledge Taighde \'{E}ireann - Research Ireland for funding Ult\'{a}n Doherty's research through a Government of Ireland Postgraduate Scholarship (GOIPG/2021/1374).}\hspace{.2cm}\\
    School of Computer Science and Statistics, Trinity College Dublin\\
    and \\
    Paul D. McNicholas \\
    Department of Mathematics and Statistics, McMaster University \\
    and \\
    Arthur White \\
    School of Computer Science and Statistics, Trinity College Dublin}
  \maketitle
} \fi

\if1\blind
{
  \bigskip
  \bigskip
  \bigskip
  \begin{center}
    {\LARGE\bf Title}
\end{center}
  \medskip
} \fi

\bigskip
\begin{abstract}
The presence of outliers can prevent clustering algorithms from accurately determining an appropriate group structure within a data set. 
We present outlierMBC, a model-based approach for sequentially removing outliers and clustering the remaining observations.
Our method identifies outliers one at a time while fitting a multivariate Gaussian mixture model to data.
Since it can be difficult to classify observations as outliers without knowing what the correct cluster structure is \emph{a priori}, and the presence of outliers interferes with the process of modelling clusters correctly, we use an iterative method to identify outliers one by one.
At each iteration, outlierMBC removes the observation with the lowest density and fits a Gaussian mixture model to the remaining data.
The method continues to remove potential outliers until a pre-set maximum number of outliers is reached, then retrospectively identifies the optimal number of outliers.
To decide how many outliers to remove, it uses the fact that the squared sample Mahalanobis distances of Gaussian distributed observations are Beta distributed when scaled appropriately.
outlierMBC chooses the number of outliers which minimises a dissimilarity between this theoretical Beta distribution and the observed distribution of the scaled squared sample Mahalanobis distances.
This means that our method both clusters the data using a Gaussian mixture model and implements a model-based procedure to identify the optimal outliers to remove without requiring the number of outliers to be pre-specified. Unlike leading methods in the literature, outlierMBC does not assume that the outliers follow a known distribution or that the number of outliers can be pre-specified.
Moreover, outlierMBC performs strongly compared to these algorithms when applied to a range of simulated and real data sets.
\end{abstract}

\noindent%
\vfill

\newpage
\spacingset{1.75} 
\section{Introduction}
\label{sec: intro}
The presence of outliers can prevent clustering algorithms from accurately determining an appropriate group structure within a data set.
This can be particularly challenging in the case of model-based clustering \citep{mcnicholas_model-based_2016, bouveyron_model-based_2019}, where departures from model assumptions can unduly influence parameter estimation and the cluster allocation of observations.
In such cases, small numbers of observations can dominate the cluster analysis, leading to an unsatisfactory analysis in which much of the underlying cluster structure of interest is missed.

We propose an iterative method for sequentially identifying outliers in a model-based clustering framework, where we assume that the mixture components are Gaussian distributions.
outlierMBC identifies and removes outliers from a data set without assuming the probability distribution of these outliers or requiring the number of outliers to be pre-specified.
Potential outliers are removed one at a time and a dissimilarity is computed at each iteration.
This dissimilarity involves the distribution of sample Mahalanobis distances.
outlierMBC presents a graph of these dissimilarity values plotted against the number of removed observations.
This dissimilarity curve is used to select the optimal number of outliers and allows the user to compare the minimum dissimilarity solution with our alternative `backtrack' solution, or choose their own number of outliers based on the graph.
outlierMBC also provides the Gaussian mixture model associated with the chosen solution, fitted to the remaining observations after the outliers have been removed.
Removing outliers one by one allows outlierMBC to methodically search for the optimal number of outliers and provides an intuitive, interpretable sequence of solutions.

\begin{figure}
	\begin{center}
		\includegraphics[width = \textwidth]{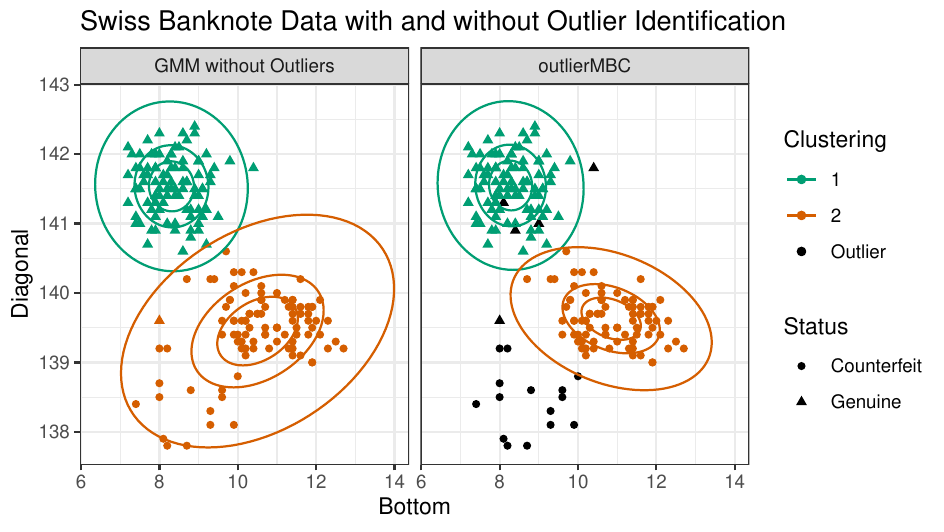}
	\end{center}
	\caption{A pair of scatter plots for two variables from the six-dimensional Swiss Banknote data set with points coloured according to clustering solutions from two-component Gaussian mixture models. The two plots compare the solution without identifying outliers (left) to the outlierMBC solution (right) which identifies twenty outliers. Density level sets, in the form of ellipses, are displayed for each component.\label{fig: banknote_contrast}}
\end{figure}

Figure \ref{fig: banknote_contrast} illustrates that removing outliers from a data set can improve the fit of a Gaussian mixture model.
It displays the Swiss Banknote data set, which will be discussed in more detail in Section \ref{sec: banknote}, clustered by a pair of two-component Gaussian mixture models, one fitted by a standard algorithm, and one fitted by outlierMBC, which identified and removed twenty outliers.
It can be observed that the second component of the standard model does not fit the counterfeit banknotes well.
However, the model fits the remaining counterfeit banknotes better after outliers are removed.
Another approach for this data set would be to fit a Gaussian mixture model with more than two components rather than removing the outliers, but this will not fully resolve the poor model fit.

Our method is implemented in R as a package called outlierMBC \citep{outliermbc}, which is published on CRAN, the Comprehensive R Archive Network. Code for reproducing this paper's analyses is available at \href{https://github.com/UltanPDoherty/outlierMBCpaper\_code}{github.com/UltanPDoherty/outlierMBCpaper\_code}.

\section{Outliers in Model-Based Clustering}
\label{subsec: MBCoutliers}

Several different approaches have been developed for dealing with outliers in model-based clustering.
Rather than explicitly differentiating between outliers and regular observations, some methods aim to accommodate points that are ill-fitting with respect to a Gaussian mixture model by replacing the Gaussian distribution with more heavy-tailed distributions such as in a mixture of $t$-distributions \citep{peel_robust_2000,evans_outlier_2015} or a mixture of power exponential distributions \citep[e.g.,][]{dang_mixtures_2015}.
Other methods use an additional component in the mixture to accommodate outliers; 
this could be a uniform component \citep{fraley_model-based_2002} or it could have an improper constant density \citep{coretto_robust_2016}.

Contaminated mixture models fit a pair of nested distributions for each component, assigning points further from the component mean, such as outliers, to the outer distribution \citep{tukey_survey_1959, punzo_parsimonious_2016}.
Another approach is to remove the outliers instead of assuming that they follow a known distribution.
\cite{garcia-escudero_general_2008} introduced a method called TCLUST for trimming a pre-specified proportion of the observations with the lowest mixture density from each iteration of its model-fitting algorithm. 

There are also two important algorithmic approaches --- called OCLUST \citep{clark_finding_2024} and OTRIMLE \citep{coretto_robust_2016}, respectively --- that use distributional results to formulate algorithms for outlier detection. To facilitate a comparison with outlierMBC, which also uses distributional results, these methods are discussed in Section~\ref{subsec: related}.

\section{Methodology}
\label{sec: method}

\subsection{outlierMBC}
In this section we outline the key details of the outlierMBC procedure. While our method operates within a model-based clustering framework, for ease of exposition we first outline the approach for a single multivariate Normal distribution. 
Consider a single multivariate Normal distribution with mean vector $\bm{\mu}$, covariance matrix $\bm{\Sigma}$, and density function $p(\cdot; \bm{\mu}, \bm{\Sigma})$.
Let $\bar{\bm{X}}$ and $\bm{S}$ be the sample mean vector and sample covariance matrix for $\bm{X}_1, \hdots, \bm{X}_n$.
Define a scaled squared sample Mahalanobis distance, $Y_i$, for each observation, $\bm{X}_i$.
The distances $Y_1, \ldots, Y_n$ are Beta-distributed \citep{gnanadesikan_robust_1972, ververidis_gaussian_2008}:
\begin{align}
	&\bm{X}_i \sim \mathcal{N}_p(\bm{\mu}, \bm{\Sigma}) \text{ for } i = 1, \hdots, n. \nonumber\\
	&Y_i := \frac{n}{(n-1)^2}\left(\bm{X}_i - \bar{\bm{X}}\right)^T \bm{S}^{-1}\left(\bm{X}_i - \bar{\bm{X}}\right) \text{ for } i = 1, \hdots, n.\nonumber\\
	&Y_i \sim \text{Beta}\left(\frac{p}{2}, \frac{n-p-1}{2}\right). \nonumber
\end{align}
In the presence of outliers, the distribution of $Y_i$ will depart from its reference Beta distribution.
outlierMBC fits a multivariate Gaussian model to the data, and then computes a dissimilarity between the observed and theoretical distributions of the scaled squared sample Mahalanobis distances.
The lowest density observation is then removed.
This process is repeated until a pre-specified maximum number of observations have been removed.
The optimal number of outliers, $o$, can then be determined based on the computed values of the dissimilarity measure.
We propose three approaches for determining $o$: 1) the value that minimises the observed-theoretical dissimilarity measure; 2) a more conservative `backtrack' solution, to be used under mild departures from our model assumptions (see Section \ref{subsec: backtrack} for further details); 3) a subjective, user-defined decision based on the graph of the dissimilarity values versus the number of removed observations.

We extend this approach to a Gaussian mixture model, defined as follows:
\begin{align*}
	&f(\bm{x}_i) = \sum_{g = 1}^G \pi_g \phi(\bm{x}_i; \bm{\mu}_g, \bm{\Sigma}_g)\\
	&\phi(\bm{x}_i; \bm{\mu}_g, \bm{\Sigma}_g) = \frac{1}{\sqrt{(2\pi)^p|\bm{\Sigma}_g|}}\exp\left\{-\frac{1}{2}(\bm{x}_i - \bm{\mu}_g)^T\bm{\Sigma}_g^{-1}(\bm{x}_i - \bm{\mu}_g)\right\}.
\end{align*}
Such models can be fitted to data using the Expectation-Maximisation (EM) algorithm \citep{dempster_maximum_1977}.
In principle, outlierMBC is agnostic to the model-fitting process, however, it does utilise the mixture R package \citep{pocuca_mixture_2024} to estimate model parameters via the EM algorithm.
In the mixture model setting, we compute a dissimilarity for each mixture component and aggregate these component-specific dissimilarities to obtain a single aggregated dissimilarity for each potential set of outliers.
We then choose the optimal number of outliers based on this aggregated dissimilarity.
Full details of the outlierMBC method are given in Algorithm \ref{alg: outlierMBC}.

\subsection{Related Methods}
\label{subsec: related}

outlierMBC and OCLUST \citep{clark_finding_2024} both remove potential outliers one by one until a user-specified maximum number of outliers have been removed and then retrospectively estimate the optimal number of outliers.
Moreover, both methods rely on the fact that scaled squared sample Mahalanobis distances are Beta-distributed \citep{ververidis_gaussian_2008, wilks_multivariate_1963}, but exploit this result in different ways.
OCLUST extends the result to show that, under certain assumptions, scaled and shifted subset log-likelihood differences are approximately Beta-distributed.
OCLUST then compares the theoretical and empirical distributions of the scaled and shifted subset log-likelihood differences as potential outliers are removed, in order to determine the optimal number of outliers in the data set.
Each subset log-likelihood difference is obtained by fitting a mixture model to the data with and without a given observation.
Therefore, if there are $n$ observations, computing $n$ subset log-likelihood differences requires the fitting of $n + 1$ Gaussian mixture models.
In contrast, outlierMBC compares the theoretical and empirical distributions of the scaled squared sample Mahalanobis distances themselves.
Since the computational cost of computing sample Mahalanobis distances is much lower than the cost of computing subset log-likelihood differences, the run time of outlierMBC is shorter than that of OCLUST.
Another benefit of our approach is that Mahalanobis distances are a more familiar and intuitive notion than subset log-likelihood differences.

OTRIMLE also uses the distribution of Mahalanobis distances as part of a model-based clustering and outlier identification algorithm.
However, there are two main differences between OTRIMLE and outlierMBC.
Firstly, OTRIMLE employs a Chi-squared distribution rather than a Beta distribution.
These are the theoretical distributions of the squared true Mahalanobis distances and of the scaled squared sample Mahalanobis distances, respectively.
The true Mahalanobis distances are computed with respect to the unknown true component mean vectors and covariance matrices, whereas the sample Mahalanobis distances are computed with respect to the sample component mean vectors and covariance matrices.
Secondly, OTRIMLE compares the theoretical and observed distributions of Mahalanobis distances to decide the improper density level of a noise component and then fits a Gaussian mixture model with this additional outlier component.
outlierMBC uses this comparison to directly choose the number of outliers.

\subsection{Sample Mahalanobis Distance Dissimilarity} \label{subsec: method_comparing}

Given a $p$-dimensional sample from a Gaussian mixture model with $G$ components of sizes $n_1, \hdots, n_G$, the theoretical distribution of the scaled squared sample Mahalanobis distances from component $g$ is a Beta distribution with parameters $\frac{p}{2}$ and $\frac{n_g-p-1}{2}$.
Using the estimated component sizes from our fitted model, $\hat{n}_1, \hdots, \hat{n}_G$, we can approximate this distribution.
We can also compute the empirical distribution of the observed scaled squared sample Mahalanobis distances with respect to our fitted model parameters for each component.

For each component, we evaluate the theoretical and empirical CDFs for the scaled squared sample Mahalanobis distances at regular intervals and compute the mean of the absolute differences between their values.
This may be seen as a sample approximation of the Wasserstein distance \citep{panaretos_statistical_2019}, $W_1(X, Y) = \int_{\mathbb{R}} |F_X(t) - F_Y(t)| dt$.
We aggregate these component-specific dissimilarity values by computing a weighted quadratic mean with respect to the component mixture proportions
Our primary choice for the optimal number of outliers is the value which minimises this aggregated dissimilarity.

\subsection{Gross Outliers} \label{subsec: method_gross}

As a pre-processing step, we identify and remove the most obvious outliers, which we call gross outliers following the terminology in \cite{clark_finding_2024}.
Our gross outlier identification procedure first computes the distance from each observation to its $k^{th}$ nearest neighbour.
By default, $k$ is set to 1\% of the sample size ($k = \lfloor 0.01 \times n \rfloor$).
We require the user to specify the maximum total number of outliers, $M$.
We identify the observations with the $M$ largest kNN distances as potential outliers.
We select the next largest kNN distance, outside of these potential outliers, as a reference value.
We classify any observation with kNN distance greater than three times the reference value to be a gross outlier.
\begin{algorithm}
	\caption{outlierMBC}
	\label{alg: outlierMBC}
	\textbf{Data, indices, outliers: } $\mathcal{X} = \{\bm{x}_1, \hdots, \bm{x}_n\} \subset \mathbb{R}^p$,\ $\mathcal{I} = \{1, \hdots, n\}$,\ $\mathcal{O} = \{\}$\\
	\textbf{User-defined constants: } $M$ = max. no. of outliers; $G$ = no. of components.\\
	\For{$m = 0$ \KwTo  $M$}{
		Let $\sigma: \{1, \hdots, n-m\} \rightarrow  \{1, \hdots, n\}$ map $\mathcal{X} \setminus \mathcal{O}$ indices to $\mathcal{X}$ indices.\\
		Fit a Gaussian mixture model to $\mathcal{X} \setminus \mathcal{O}$, estimating the parameters, $\bm{\hat{\Psi}} = \{\hat{\pi}_g,\ \bm{\hat{\mu}_g},\ \bm{\hat{\Sigma}_g}\}_{g = 1}^{G}$, and the component assignment probability matrix, $\hat{Z}$.
		
		\For{$g = 1$  \KwTo $G$}{
			Estimate the number of data points: \hfill $\hat{n}_g \gets \sum_{j = 1}^{n-m} \hat{Z}_{jg}$.
			
			Compute the sample covariance matrix: \hfill $\bm{\hat{S}}_g \gets \frac{\hat{n}_g}{\hat{n}_g - 1}\bm{\hat{\Sigma}}_g$.
			
			Compute the squared sample Mahalanobis distances:\\
			\hfil $\hat{d}_g(\bm{x_{\sigma(j)}})^2 = (\bm{x_{\sigma(j)}} - \bm{\hat{\mu}}_g)^T\bm{\hat{S}}_g^{-1}(\bm{x_{\sigma(j)}} - \bm{\hat{\mu}}_g)$
			
			Rescale the squared sample Mahalanobis distances: \hfill $\hat{y}_{jg} \gets \frac{\hat{n}_g}{(\hat{n}_g - 1)^2}\hat{d}_g(\bm{x_{\sigma(j)}})^2$.
			
			Let $\hat{F}_g$ be the empirical CDF of $\{\hat{y}_{jg}\}_{j = 1}^{n-m}$ weighted by $\left\{\frac{\hat{Z}_{jg}}{\hat{n}_g}\right\}_{j = 1}^{n-m}$.
			
			Let $F_g$ be the CDF of a Beta distribution with parameters $\frac{p}{2}$ and $\frac{\hat{n}_g-p-1}{2}$.
			
			Compute the mean of the absolute CDF differences ($T = 10,000$):\\
			\hfil $\hat{D}_{mg} \gets \frac{1}{T}\sum_{t=1}^{T} \left|F_g\left(\frac{t}{T}\right) - \hat{F}_g\left(\frac{t}{T}\right)\right|$ \\
		}
		
		Aggregate the component-wise dissimilarities: \hfill $\hat{D}_m \gets \sqrt{\sum_{g=1}^G\hat{\pi}_g\hat{D}_{mg}^2}$.\\
		
		Identify the lowest mixture density data point: \hfill $\text{out}(m) \gets ^{\arg \min}_{\ \ i \in \mathcal{I}} \left\{ p(\bm{x}_i; \bm{\hat{\Psi}})\right\}$.
		
		Add it to the set of outliers: \hfill $\mathcal{O} \gets \mathcal{O} \cup \{\bm{x}_{\text{out}(m)}\}$.
		
		Remove it from the current set of indices: \hfill $\mathcal{I} \gets \mathcal{I} \setminus \{\text{out}(m)\}$.
	}
	
	Identify when the aggregated dissimilarity was minimised: \hfill $o \gets ^{\ \ \arg \min}_{m \in \{0,\hdots,M\}}\left\{\hat{D}_m\right\}$.\\
	
	Classify the first $o$ points removed as outliers: \hfill $\mathcal{O} \gets \{\bm{x}_{\text{out}(1)}, \bm{x}_{\text{out}(2)}, \hdots, \bm{x}_{\text{out}(o)}\}$.
	
	Fit a $G$-component Gaussian mixture model to $\mathcal{X} \setminus \mathcal{O}$.\\
\end{algorithm}

\subsection{backtrack Mechanism} \label{subsec: backtrack}

In some situations, particularly when the data does not fully satisfy the assumptions of the Gaussian mixture model, the solution which exactly minimises the sample Mahalanobis distance dissimilarity may remove more observations than desired.
For these cases, we have developed a mechanism which allows outlierMBC to select a more conservative number of outliers by moving backwards from the minimum.
Each backward step reduces the chosen number of outliers by one and must pass two validity checks in order to be allowed.
These checks are defined based on two user-defined thresholds, namely, the maximum step rise and the maximum total rise, which we will refer to as $\alpha$ and $\beta$, respectively.
Firstly, the increase in the dissimilarity as a result of a single step, as a proportion of the minimum dissimilarity, must be less than $\alpha$.
Secondly, the cumulative increase from all steps cannot exceed $\beta$, as a proportion of the minimum dissimilarity.
Both of these parameters can be specified by the user, however the defaults are $\alpha = 0.05$ and $\beta = 0.10$, meaning that the dissimilarity cannot increase by more than 5\% of the minimum in any one step and cannot increase by more than 10\% of the minimum in total.
We will refer to the solutions obtained by choosing the exact minimum as outlierMBC-minimum and the solution which uses the backtrack choice as outlierMBC-backtrack.

Similarly, OTRIMLE has a procedure which allows it to accept a choice of its improper density value which is sub-optimal with respect to minimising its dissimilarity but results in a lower number of outliers being removed.
It does so by minimising a penalised version of the dissimilarity.
The penalty term consists of the proportion of outliers multiplied by a user-defined quantity $\beta$.

\subsection{Initialisation} \label{subsec: initialisation}

outlierMBC uses a hierarchical clustering approach to construct an initial cluster solution to initialise the first Gaussian mixture model.
However, since outlierMBC fits a large sequence of Gaussian mixture models, we designed two different procedures for how to initialise the remaining models in the sequence.
Our first initialisation scheme, which we will refer to as the `update' scheme, takes the posterior component assignment probability matrix from the fitted previous model, removes the row corresponding to the newly proposed outlier, and then uses this submatrix to initialise the next mixture model.
Our second initialisation scheme, which we will refer to as `reinit', implements hierarchical clustering after every outlier removal and provides a brand new reinitialisation to the next model.
The `update' scheme is usually much quicker and for many data sets, the extra reinitialisation can be redundant.
However, the `reinit' scheme can be more suitable for particularly challenging data sets where there is no clear cluster structure.

\section{Illustrative Examples} \label{sec: ill_examp}

\begin{figure}
	\begin{center}
		\includegraphics[width = \textwidth]{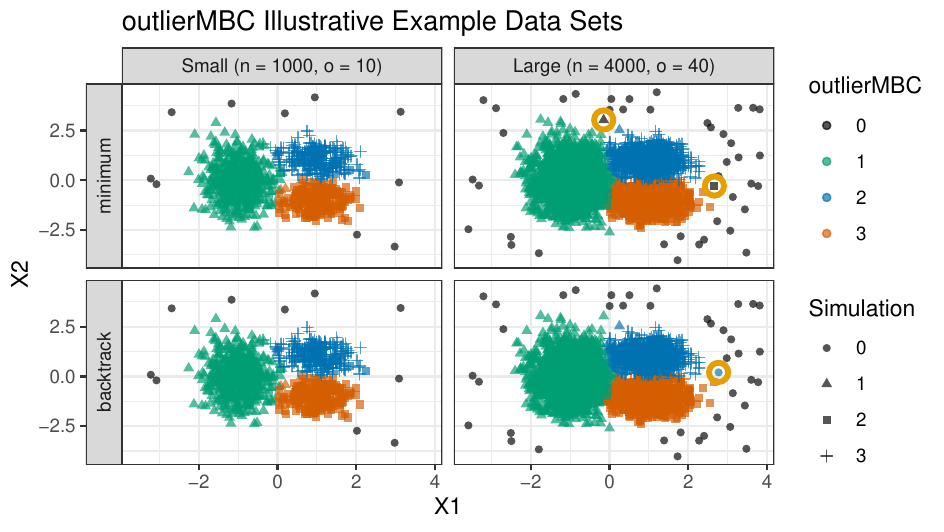}
	\end{center}
	\caption{Scatter plots of the illustrative data sets with their points coloured according to the outlierMBC solutions and with outlier classification errors circled in orange.\label{fig: packgmm_scatter}}
\end{figure}

\begin{figure}
	\begin{center}
		\includegraphics[width = \textwidth]{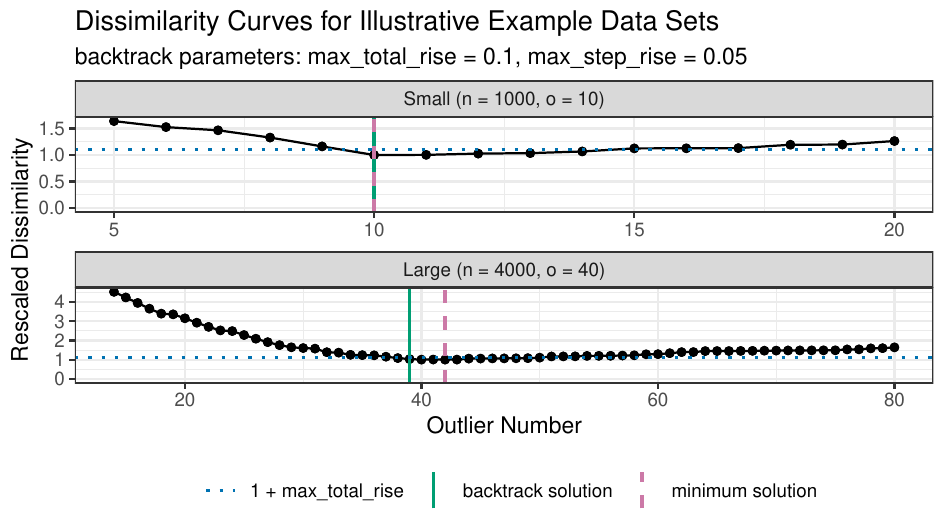}
	\end{center}
	\caption{Dissimilarity curves for the two illustrative data sets. The dissimilarity values on the y axis have been rescaled so that the minimum has a value of 1.\label{fig: packgmm_backtrack_curves}}
\end{figure}

To demonstrate how the method works in relatively straightforward conditions, we applied it to small and large data sets simulated from the same three-component mixture model.
The mean vectors, covariance matrices, and mixture proportions of the three components are the same for both data sets.
The data sets consist of 1000 or 4000 Gaussian observations and 10 or 40 outliers.
The true observations were sampled from a mixture of Gaussian distributions and the outliers were sampled from a Uniform distribution across a hyper-rectangle containing the true observations excluding an ellipsoid around each Gaussian distribution.
These data sets are shown in Figure \ref{fig: packgmm_scatter} with observations coloured according to their outlierMBC clustering.
outlierMBC-minimum and outlierMBC-backtrack both achieved perfect classification for the small data set while for the large data set, they had two false positives and one false negative, respectively.
The dissimilarity curves for both data sets are smooth and stable, as can be seen in Figure \ref{fig: packgmm_backtrack_curves}.

\section{TCLUST Simulation} \label{sec: tclust_sim}

\subsection{TCLUST Simulation Setup} \label{subsec: tclust_sim_setup}

\begin{figure}
	\begin{center}
		\includegraphics[width = \textwidth]{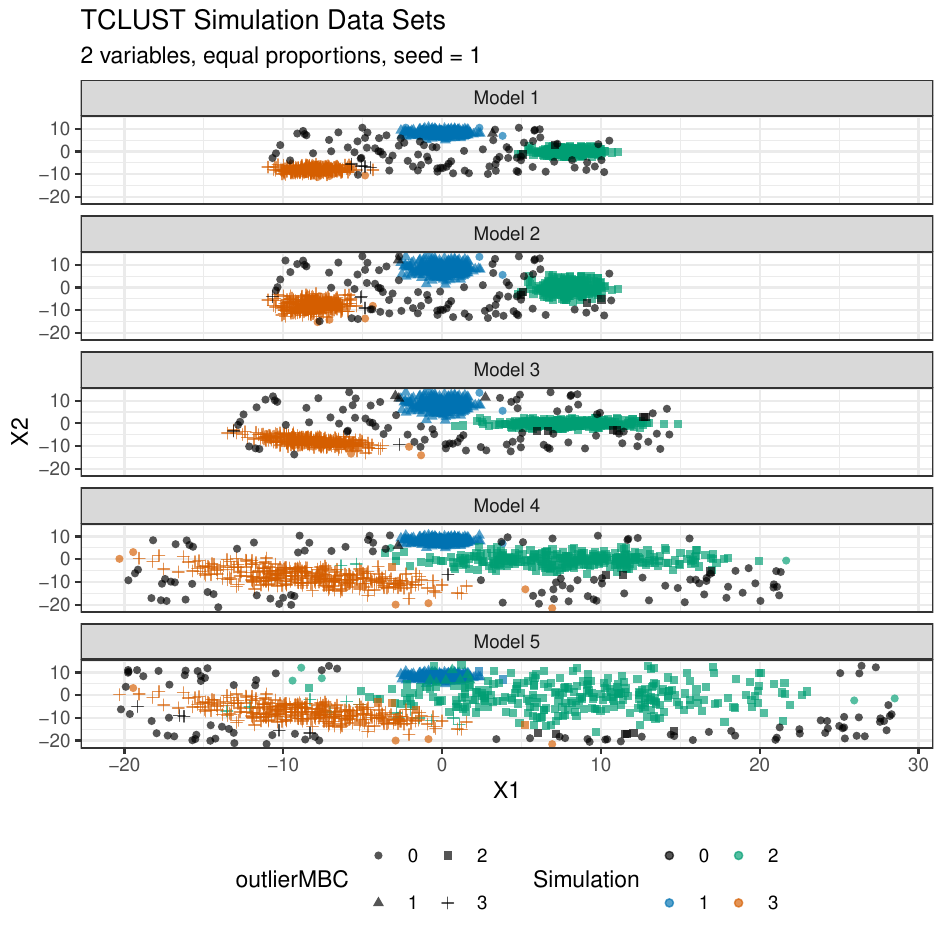}
	\end{center}
	\caption{Scatter plots for the two-dimensional models (seed = 1) with equal component sizes. Observations' colours correspond to their simulation source and their shapes correspond to the outlierMBC-minimum classification.\label{fig: ombc_p2n1s1_scatter}}
\end{figure}

This simulation study was designed to mimic one from \cite{garcia-escudero_general_2008}.
The \texttt{simulate\_gmm} function from the outlierMBC R package was used to simulate 200 data sets, each consisting of 900 observations from three Gaussian components and 100 outliers. Table \ref{tab: tclust_options} shows the settings which were varied to obtain 200 data sets from 20 different simulation scenarios. Tables \ref{tab: tclust_model_params} and \ref{tab: tclust_abcdef} describe the mean vectors and covariance matrices for the mixture components. Figure \ref{fig: ombc_p2n1s1_scatter} displays the five simulated data sets that have two variables, equal mixture proportions, and a random seed of 1.

To simulate a data set, we began by sampling 300 observations from each Gaussian component.
Then, to simulate an outlier, a sample was drawn from a uniform distribution over the hyper-rectangle containing the Gaussian data.
This sample was only accepted if it was extreme with respect to every component in the mixture, in particular, if it was outside that component's theoretical 99$^{th}$ percentile ellipsoid.
To determine this, we calculated the sample's Mahalanobis distance from that component's true mean using its true covariance matrix and then checked if it was greater than the 99$^{th}$ percentile of a Chi-squared distribution with $p$ degrees of freedom where $p$ is the dimension.
Uniform samples were simulated until 100 were accepted as outliers.
\begin{table}[p]
	\caption{Simulation variable options.\label{tab: tclust_options}}
	\begin{center}
		\begin{tabular}{llc}
			\toprule
			Simulation Variable & Options & Number \\
			\midrule
			Dimension, $p$ & $2$ or $6$ & 2 \\
			Component Proportions & $\left(\frac{1}{3},\ \frac{1}{3},\ \frac{1}{3}\right)$ or $\left(\frac{1}{5},\ \frac{2}{5},\ \frac{2}{5}\right)$ & 2 \\
			Covariance Constants & Model 1, 2, 3, 4, or 5 & 5 \\
			Random Seeds & 1, 2, 3, 4, 5, 6, 7, 8, 9, or 10 & 10 \\
			\bottomrule
		\end{tabular}
	\end{center}
\end{table}
\begin{table}[p]
	\caption{Component mean vectors and covariance matrices for the TCLUST simulation study. The values of the covariance constants $a$, $b$, $c$, $d$, $e$, and $f$ are provided in Table \ref{tab: tclust_abcdef}.\label{tab: tclust_model_params}}
	\begin{center}
		\begin{tabular}{ll}
			\toprule
			$\bm{\mu}_g^{(p)}$ & $\bm{\Sigma}_g^{(p)}$ \\
			\midrule
			$\bm{\mu}_1^{(2)} = (0, 8)$ & $\bm{\Sigma}_1^{(2)}$ = $\begin{pmatrix}
				1 & 0 \\
				0 & a
			\end{pmatrix}$ \\
			$\bm{\mu}_2^{(2)} = (8, 0)$ & $\bm{\Sigma}_2^{(2)} = \begin{pmatrix}
				b & 0 \\
				0 & c
			\end{pmatrix}$ \\
			$\bm{\mu}_3^{(2)} = (-8, -8)$ & $\bm{\Sigma}_3^{(2)} = \begin{pmatrix}
				d & e \\
				e & f
			\end{pmatrix}$ \\
			\bottomrule
		\end{tabular}\hfil\begin{tabular}{ll}
			\toprule
			$\bm{\mu}_g^{(p)}$ & $\bm{\Sigma}_g^{(p)}$ \\
			\midrule
			$\bm{\mu}_1^{(6)} = \left(\bm{\mu}_1^{(2)}, 0, 0, 0, 0\right)$ & $\bm{\Sigma}_1^{(6)} = \begin{pmatrix}
				\bm{\Sigma}_1^{(2)} & 0 \\
				0 & \bm{I}_4
			\end{pmatrix}$ \\
			$\bm{\mu}_2^{(6)} = \left(\bm{\mu}_2^{(2)}, 0, 0, 0, 0\right)$ & $\bm{\Sigma}_2^{(6)} = \begin{pmatrix}
				\bm{\Sigma}_2^{(2)} & 0 \\
				0 & \bm{I}_4
			\end{pmatrix}$ \\
			$\bm{\mu}_3^{(6)} = \left(\bm{\mu}_3^{(2)}, 0, 0, 0, 0\right)$ & $\bm{\Sigma}_3^{(6)} = \begin{pmatrix}
				\bm{\Sigma}_3^{(2)} & 0 \\
				0 & \bm{I}_4
			\end{pmatrix}$ \\
			\bottomrule
		\end{tabular}
	\end{center}
\end{table}
\begin{table}[p]
	\caption{Covariance constants, $(a,\ b,\ c,\ d,\ e,\ f)$, for the TCLUST simulation study.\label{tab: tclust_abcdef}}
	\begin{center}
		\begin{tabular}{ccc}
			\toprule
			Model 1 & Model 2 & Model 3 \\
			\midrule
			(1, 1, 1, 1, 0, 1) & (5, 1, 5, 1, 0, 5) & (5, 5, 1, 3, -2, 3) \\
			\midrule
		\end{tabular}
		\begin{tabular}{cc}
			Model 4 & Model 5 \\
			\midrule
			(1, 20, 5, 15, -10, 15) & (1, 45, 30, 15, -10, 15) \\
			\bottomrule
		\end{tabular}
	\end{center}
\end{table}

\subsection{TCLUST Simulation Results} \label{subsec: tclust_sim_results}

We applied outlierMBC-minimum and outlierMBC-backtrack to these data sets, as well as a range of leading model-based methods for clustering and outlier identification.
These methods were TCLUST \citep{garcia-escudero_general_2008}, ContaminatedMixt, a contaminated Gaussian mixture \citep{punzo_contaminatedmixt_2018}, mclust, a Gaussian mixture with a Uniform component \citep{fraley_model-based_2002, fraley_mclust_2022}, and OTRIMLE \citep{coretto_robust_2016}.
We initialised mclust following the approach proposed by \cite{scrucca_entropy-based_2023}.

ContaminatedMixt fits both contaminated and uncontaminated models but for 20 of the 200 data sets in this simulation study, its best model according to BIC (Bayesian Information Criterion, \cite{schwarz_estimating_1978}) was uncontaminated. This means that for these data sets, ContaminatedMixt did not classify any data points as outliers. Similarly, OTRIMLE did not identify any outliers for 14 of the 200 data sets.

Table \ref{tab: tclust_mean_vals} shows that outlierMBC-minimum was the top performing method in terms of both mean F1 and mean ARI, despite TCLUST being provided with the true number of simulated outliers.
outlierMBC-minimum also identified fewer false positives on average than TCLUST, mclust, OTRIMLE, and ContaminatedMixt.
outlierMBC-backtrack offered a more conservative alternative with fewer false positives but more false negatives.
\begin{table}
	\caption{Mean values for the TCLUST simulation study of Adjusted Rand Index (ARI), F1 Measure with respect to outlier classification only (Outlier F1), and Numbers of False Positives (\#FP), False Negatives (\#FN), and Outliers (\#Outliers).\label{tab: tclust_mean_vals}}
	\begin{center}
		\begin{tabular}{lccccc}
			\toprule
			Method & ARI & Outlier F1 & \#FP & \#FN & \#Outliers \\
			\midrule
			outlierMBC-minimum   & 0.96 & 0.94 & 4.50 & 8.00 & 96.50 \\
			outlierMBC-backtrack & 0.96 & 0.92 & 3.14 & 11.75 & 91.39 \\
			TCLUST ($\alpha = 0.1$) & 0.95 & 0.93 & 6.76 & 6.76 & 100 \\
			mclust               & 0.94 & 0.90 & 25.32 & 1.02 & 124.30 \\
			ContaminatedMixt     & 0.89 & 0.77 & 32.99 & 14.26 & 118.74 \\
			OTRIMLE              & 0.86 & 0.83 & 10.13 & 16.18 & 93.96 \\
			\bottomrule
		\end{tabular}
	\end{center}
\end{table}

ContaminatedMixt, mclust, and OTRIMLE failed quite dramatically for some of the data sets, whereas outlierMBC was more reliable.
The highest number of false positives identified by ContaminatedMixt was 295, compared to 204 for mclust, 67 for OTRIMLE, 22 for TCLUST, 14 for outlierMBC-minimum, and 13 for outlierMBC-backtrack.
ContaminatedMixt and mclust had more than 50 false positives for 37 data sets and 25 data sets, respectively, while OTRIMLE only exceeded this number once.
ContaminatedMixt and OTRIMLE exceeded 50 false negatives 22 times and 25 times, respectively, compared to twice for outlierMBC-backtrack and once for outlierMBC-minimum.
This shows that outlierMBC was less likely to produce an extreme number of false positives or false negatives.

\section{OCLUST Simulation} \label{sec: oclust_sim}

For this simulation study, we replicated eight data sets from \cite{clark_finding_2024} by adding outliers to data sets from \cite{franti_k-means_2018}: s1 - s4, a1 - a3, and Unbalance (\href{https://cs.uef.fi/sipu/datasets}{cs.uef.fi/sipu/datasets}).
The outliers were sampled uniformly across a hyper-rectangle centred at the mean of each data set with twice the range in each dimension as the data.
There was no rejection procedure, meaning that some of these uniform samples lie within the clusters and some false negatives are to be expected when identifying outliers.
When applying outlierMBC, we used the `reinit' initialisation scheme for the s3 and s4 data sets, and the default `update' scheme for the other six data sets.

Both outlierMBC and OCLUST implement a preliminary gross outlier removal step. To facilitate a direct comparison and to avoid the subjective elbow choice required by OCLUST's procedure, we implemented both methods with outlierMBC's gross outliers.

From Tables \ref{tab: oclustsim_ari} and \ref{tab: oclustsim_f1}, we can see that outlierMBC-backtrack achieves the highest mean ARI and mean F1 values across these eight data sets.
Moreover, Table \ref{tab: oclustsim_fp} shows that it achieves this while identifying fewer false positives on average than the other methods.
\begin{table}
	\caption{ARI values from the OCLUST simulation study.\label{tab: oclustsim_ari}}
	\begin{center}
		\begin{tabular}{lcccccccc|c}
			\toprule
			Method & a1 & a2 & a3 & s1 & s2 & s3 & s4 & Unbal. & Mean \\
			\midrule
			outlierMBC-minimum   & 0.95 & 0.94 & 0.93 & 0.92 & 0.88 & 0.68 & 0.52 & 1.00 & 0.85 \\
			outlierMBC-backtrack & 0.95 & 0.94 & 0.94 & 0.96 & 0.90 & 0.71 & 0.52 & 1.00 & 0.87 \\
			OCLUST               & 0.96 & 0.94 & 0.93 & 0.96 & 0.90 & 0.59 & 0.41 & 1.00 &   0.84 \\
			mclust               & 0.97 & 0.90 & 0.66 & 0.96 & 0.89 & 0.71 & 0.57 & 0.98 & 0.83 \\
			OTRIMLE              & 0.94 & 0.93 & 0.92 & 0.93 & 0.89 & 0.54 & 0.58 & 1.00 & 0.84 \\
			ContaminatedMixt     & 0.80 & 0.63 & 0.47 & 0.87 & 0.13 & 0.54 & 0.22 & 0.96 & 0.58 \\
			\bottomrule
		\end{tabular}
	\end{center}
\end{table}
\begin{table}
	\caption{Outlier F1 values from the OCLUST simulation study.\label{tab: oclustsim_f1}}
	\begin{center}
		\begin{tabular}{lcccccccc|c}
			\toprule
			Method & a1 & a2 & a3 & s1 & s2 & s3 & s4 & Unbal. & Mean \\
			\midrule
			outlierMBC-minimum   & 0.90 & 0.87 & 0.86 & 0.77 & 0.77 & 0.77 & 0.88 & 0.97 & 0.85 \\
			outlierMBC-backtrack & 0.90 & 0.88 & 0.88 & 0.88 & 0.86 & 0.87 & 0.88 & 0.97 & 0.89 \\
			OCLUST               & 0.92 & 0.88 & 0.87 & 0.89 & 0.86 & 0.84 & 0.81 & 0.97 &   0.88 \\
			mclust               & 0.92 & 0.87 & 0.85 & 0.87 & 0.87 & 0.85 & 0.88 & 0.94 & 0.88 \\
			OTRIMLE              & 0.88 & 0.85 & 0.85 & 0.80 & 0.80 & 0.00 & 0.82 & 0.94 & 0.74 \\
			ContaminatedMixt     & 0.00 & 0.00 & 0.00 & 0.73 & 0.23 & 0.00 & 0.00 & 0.00 & 0.12 \\
			\bottomrule
		\end{tabular}
	\end{center}
\end{table}
\begin{table}
	\caption{Number of false positives from the OCLUST simulation study.\label{tab: oclustsim_fp}}
	\begin{center}
		\begin{tabular}{lcccccccc|c}
			\toprule
			Method & a1 & a2 & a3 & s1 & s2 & s3 & s4 & Unbal. & Mean \\
			\midrule
			outlierMBC-minimum   & 16 & 38 & 55 & 142 & 135 & 111 &   7 & 12 & 64 \\
			outlierMBC-backtrack &  0 &  2 &  5 &  17 &   0 &   5 &   7 &  8 &  6 \\
			OCLUST               &  0 &  0 &  1 &   3 &   1 &  22 & 103 &  8 & 17 \\
			mclust               &  1 & 13 &  6 &  29 &  17 &  25 &  25 &  2 & 15 \\
			OTRIMLE              & 28 & 54 & 83 & 116 &  96 &   0 &  89 & 49 & 64 \\
			ContaminatedMixt     &  0 &  0 &  0 & 163 & 2202 &  0 & 0 & 0 & 296 \\
			\bottomrule
		\end{tabular}
	\end{center}
\end{table}

outlierMBC-backtrack outperforms outlierMBC-minimum for these data sets.
By design, the number of outliers selected by outlierMBC-backtrack must be less than or equal to the number selected by outlierMBC-minimum.
outlierMBC-minimum often identified a high number of outliers relative to the other methods and from Table \ref{tab: oclustsim_fp} we can see that this resulted in a high number of false positives.
By identifying fewer outliers, outlierMBC-backtrack was able to avoid these false positives.
It also matched or outperformed outlierMBC-minimum with respect to ARI and F1 for every data set, while selecting fewer outliers.
Since the F1 measure is the harmonic mean of recall and precision, this suggests that outlierMBC-backtrack more than made up for any decrease in its outlier classification recall with an increase in precision.
Furthermore, its high ARI score indicates that choosing not to remove these observations did not prevent outlierMBC-backtrack from uncovering the true cluster structure of these data sets.

\begin{figure}
	\begin{center}
		\includegraphics[width = \textwidth]{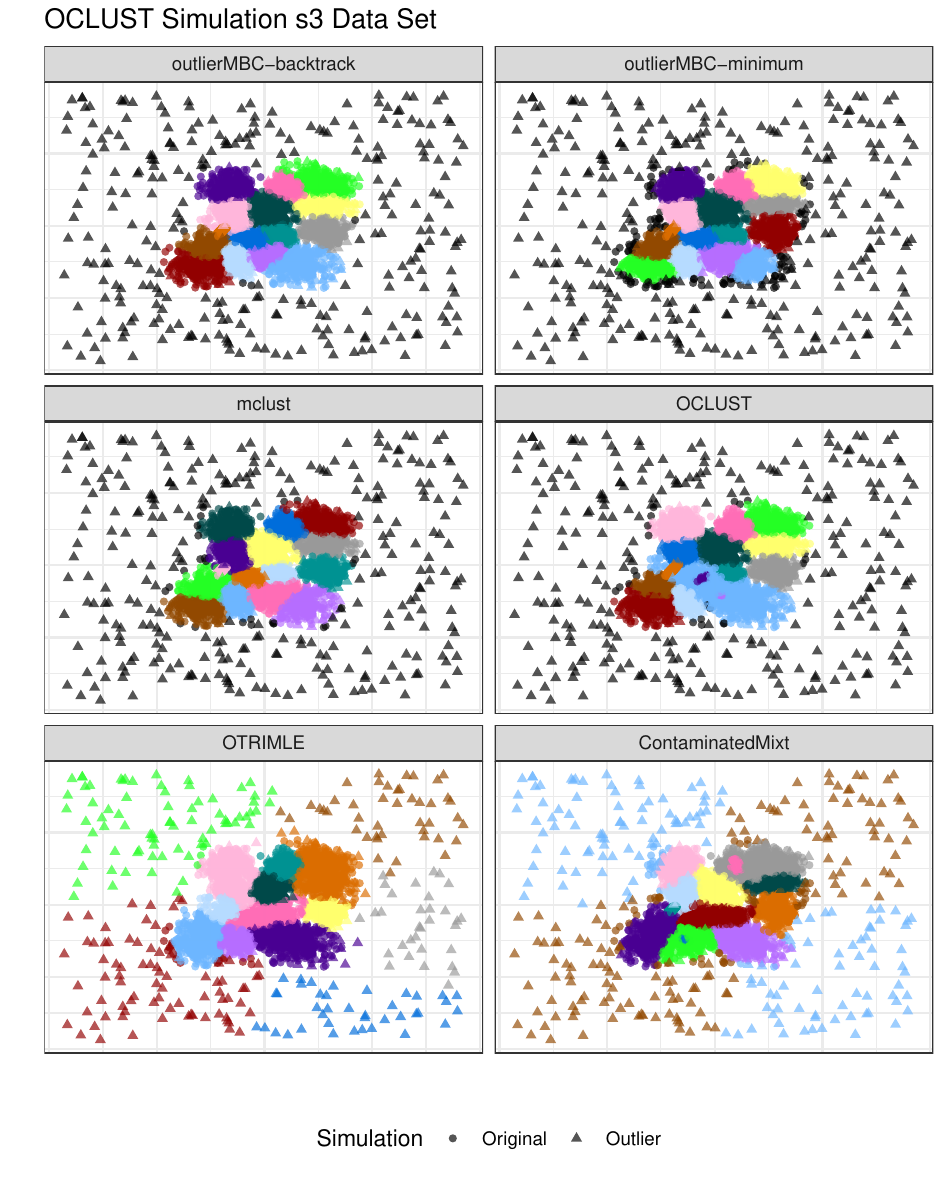}
	\end{center}
	\caption{Clustering and outlier identification solutions for the s3 data set.\label{fig: s3_scatters}}
\end{figure}

Figure \ref{fig: s3_scatters} displays six scatter plots of the s3 data set each coloured according to the clustering solution obtained by one of the methods applied in this simulation study.
Notably, the OTRIMLE and ContaminatedMixt solutions do not identify any outliers in this case.
From Tables \ref{tab: oclustsim_ari}, \ref{tab: oclustsim_f1}, and \ref{tab: oclustsim_fp}, we can observe that outlierMBC-backtrack had the joint highest ARI (0.71) and the highest Outlier F1 (0.87) for this data set.
Among the methods that identified any outliers, it also had the lowest number of false positives (5).

\section{Swiss Banknote Data} \label{sec: banknote}

The Swiss Banknote data set consists of 200 observations of 6 numerical variables corresponding to a set of measurements for 100 genuine and 100 counterfeit banknotes.
In the TCLUST R package paper \citep{fritz_tclust_2012}, the authors propose using TCLUST for this data set with 10\% of observations classified as outliers.
outlierMBC obtains the exact same solution without having to pre-specify this proportion.

Table \ref{tab: banknote_comparison} shows a comparison of the clustering results for a number of leading methods applied to the Swiss Banknote data set. outlierMBC-minimum, outlierMBC-backtrack, TCLUST ($\alpha = 0.1$), and OCLUST reach an identical clustering solution, which is displayed in the right-hand plot of Figure \ref{fig: banknote_contrast}.
mclust and OTRIMLE differ by one point each with mclust removing an additional counterfeit banknote and OTRIMLE retaining a genuine banknote.
ContaminatedMixt does not classify any of the banknotes as outliers but does assign 17 counterfeit banknotes to the same cluster as the 100 genuine banknotes.
In contrast, none of the other methods have both genuine and counterfeit banknotes assigned to the same component.
\begin{table}
	\caption{These tables show each method's cluster assignment of the genuine banknotes (left) and the counterfeit banknotes (right) from the Swiss Banknote data set.\label{tab: banknote_comparison}}
	\begin{center}
		\begin{tabular}{lccc}
			\toprule
			True Class & \multicolumn{3}{c}{Genuine} \\
			Assigned Cluster & 1 & 2 & Outlier \\
			\midrule
			outlierMBC-minimum      &  95 & 0 & 5 \\
			outlierMBC-backtrack    &  95 & 0 & 5 \\
			TCLUST ($\alpha = 0.1$) &  95 & 0 & 5 \\
			OCLUST    			    &  95 & 0 & 5 \\
			mclust    			    &  95 & 0 & 5 \\
			OTRIMLE    			    &  96 & 0 & 4 \\
			ContaminatedMixt        & 100 & 0 & 0 \\
			\bottomrule
		\end{tabular} \hfill\begin{tabular}{lccc}
			\toprule
			True Class & \multicolumn{3}{c}{Counterfeit} \\
			Assigned Cluster & 1 & 2 & Outlier \\
			\midrule
			outlierMBC-minimum      &  0 & 85 & 15 \\
			outlierMBC-backtrack    &  0 & 85 & 15 \\
			TCLUST ($\alpha = 0.1$) & 0 & 85 & 15 \\
			OCLUST    			    &  0 & 85 & 15 \\
			mclust    			    &  0 & 84 & 16 \\
			OTRIMLE    			    &  0 & 85 & 15 \\
			ContaminatedMixt        & 17 & 83 & 0 \\
			\bottomrule
		\end{tabular}
	\end{center}
\end{table}

\section{Conclusion}

We have presented a method for sequentially identifying outliers by comparing the observed distribution of the data's scaled squared sample Mahalanobis distances to their theoretical distribution.
We have replicated simulation studies from \cite{garcia-escudero_general_2008} and \cite{clark_finding_2024} and compared outlierMBC to leading methods for model-based clustering and outlier identification.
These included TCLUST, OCLUST, mclust, and OTRIMLE.
outlierMBC consistently performed well with respect to replicating the true cluster structure (ARI) and classifying outliers (Outlier F1).
It also demonstrated a tendency to identify fewer false positives than the other methods.
This is an appealing characteristic for an outlier identification method as unnecessarily removing true observations could result in a misleading view of the data.

One area of future work which we have been exploring in relation to this method is extending outlierMBC to Gaussian linear cluster-weighted models rather than Gaussian mixture models, by exploiting the Beta distribution of Studentised residuals. This work is in its early stages but has delivered promising preliminary results. Another possible extension would be to non-Gaussian mixture models. This would require identifying a non-Gaussian analogue to the Mahalanobis distance and determining its theoretical distribution. For example, it could be interesting to examine how our outlier removal procedure would operate with respect to the more robust, heavy-tailed distributions of a $t$ mixture or the more flexible distributions of a skew-Normal mixture.

\bibliographystyle{chicago}

\end{document}